\documentclass[conference]{IEEEtran}
\IEEEoverridecommandlockouts
\usepackage[colorlinks,citecolor=blue,urlcolor=blue,bookmarks=false,hypertexnames=true]{hyperref} 
\usepackage{cite}
\usepackage{amsmath,amssymb,amsfonts}
\usepackage{graphicx}
\usepackage{textcomp}
\usepackage{svg}
\usepackage{float}
\usepackage{amsmath}
\usepackage{multicol}
\usepackage{array}
\usepackage[noend]{algpseudocode}
\usepackage[ruled, lined, linesnumbered, commentsnumbered, longend]{algorithm2e}
\usepackage{multirow}
\usepackage{adjustbox}

\usepackage{colortbl}
\usepackage{graphicx}

\definecolor{blue1}{HTML}{CBD4DF}
\definecolor{blue2}{HTML}{A0B4C7}
\definecolor{blue3}{HTML}{7998B3}
\definecolor{blue4}{HTML}{5181A1}
\definecolor{blue5}{HTML}{156E92}
\definecolor{blue6}{HTML}{486B84}
\definecolor{green1}{HTML}{D9EDDC}
\definecolor{green2}{HTML}{B6DDBE}
\definecolor{green3}{HTML}{93CFA1}
\definecolor{green4}{HTML}{6DC288}
\definecolor{green5}{HTML}{3CB76D}
\definecolor{green6}{HTML}{AEDBC2}
\definecolor{green7}{HTML}{C2E3D0}
\definecolor{green8}{HTML}{D5ECDE}

\definecolor{green9}{HTML}{E4F3EA}
\definecolor{greens}{HTML}{ECF6EF}
\definecolor{greens1}{HTML}{FAFDFA}

\definecolor{greens2}{HTML}{BCE2CE}
\definecolor{greens3}{HTML}{CDE8D9}
\definecolor{greens5}{HTML}{ECF6F0}
\definecolor{gold1}{HTML}{EFDFC8}
\definecolor{gold2}{HTML}{E3C69C}
\definecolor{gold3}{HTML}{D84F74}
\definecolor{gold4}{HTML}{CD994D}
\definecolor{gold5}{HTML}{C3862B}

\definecolor{golds}{HTML}{F1E4D1}
\definecolor{golds1}{HTML}{F6EDE1}

\definecolor{gold_fig}{HTML}{F8D287}
\definecolor{green_fig}{HTML}{EBF5EC}

\usepackage{lipsum}

\def\BibTeX{{\rm B\kern-.05em{\sc i\kern-.025em b}\kern-.08em
    T\kern-.1667em\lower.7ex\hbox{E}\kern-.125emX}}
    

\begin{document}

\title{Support for Stock Trend Prediction Using Transformers and Sentiment Analysis \\
}

\author{
\IEEEauthorblockN{Harsimrat Kaeley}

\IEEEauthorblockA{\textit{Department of Electrical Engineering}\\{and Computer Science}\\
\textit{University of California, Irvine}\\
Irvine, California, USA\\
kaeleyh@uci.edu}
\and
\IEEEauthorblockN{Ye Qiao}
\IEEEauthorblockA{\textit{Department of Electrical Engineering}\\\textit{and Computer Science}\\
\textit{University of California, Irvine}\\
Irvine, California, USA\\
yeq6@uci.edu}
\and
\IEEEauthorblockN{Nader Bagherzadeh, \textit{Fellow, IEEE}}
\IEEEauthorblockA{\textit{Department of Electrical Engineering}\\\textit{and Computer Science}\\
\textit{University of California, Irvine}\\
Irvine, California, USA\\
nader@uci.edu}
}

\maketitle
\begin{sloppypar}

\begin{abstract}

Stock trend analysis has been an influential time-series prediction topic due to its lucrative and inherently chaotic nature. Many models looking to accurately predict the trend of stocks have been based on Recurrent Neural Networks (RNNs). However, due to the limitations of RNNs, such as gradient vanish and long-term dependencies being lost as sequence length increases, in this paper we develop a Transformer based model that uses technical stock data and sentiment analysis to conduct accurate stock trend prediction over long time windows. This paper also introduces a novel dataset containing daily technical stock data and top news headline data spanning almost three years. Stock prediction based solely on technical data can suffer from lag caused by the inability of stock indicators to effectively factor in breaking market news. The use of sentiment analysis on top headlines can help account for unforeseen shifts in market conditions caused by news coverage. We measure the performance of our model against RNNs over sequence lengths spanning 5 business days to 30 business days to mimic different length trading strategies. This reveals an improvement in directional accuracy over RNNs as sequence length is increased, with the largest improvement being close to 18.63\% at 30 business days.

\end{abstract}


\section{Introduction}
Investing in the stock market can prove to be a very lucrative endeavor, to such an extent that massive corporations exist today solely by gleaning most of their profit from predicting stock market trends. Consequently, tools that help make investment decisions are of high value, and the onset of new markets bolster their demand. 

Deep convolutional neural networks (CNN) have achieved incredible results in computer vision, speech recognition, object detection, natural language processing, and many other fields \cite{lecun2015deep} \cite{qiao2022two} \cite{ding2023bnn}.
Deep learning is unlike any other market prediction tactic in that it allows for the creation of multi-layered computational models. Given a dataset that contains input features representing different aspects of the prediction problem, these models are able to extract latent relationships between the features. Each layer builds upon the previous layer's feature output to learn different levels of abstraction, but when combined they accurately represent the prediction problem \cite{lecun2015deep}.

The prediction task in this paper consists of using technical and sentiment data from an \emph{n} day lag period to predict the (\emph{n}+1) day's normalized opening price. We normalized each technical feature and label in our novel dataset to allow our model to focus on trend prediction over the specifics of stock pricing.

Our model relies on two factors for its performance increase over increased sequence lengths compared to conventional RNNs. The first being its Transformer backbone's ability to effectively capture long-term dependencies, through the use of self-attention \cite{attention}. By using self-attention, the context between each day in a given input sequence is maintained as it is directly encoded into the input. 

The second factor offering an increase in performance is the extra context offered by conducting sentiment analysis on news headlines. Predicting stock trends using only technical data analysis does well at capturing overall market conditions, but its inherent lag and use of past estimates can lead to a disconnect between the data and current stock movements \cite{lawrence1997using}. Therefore the latest news, which is most relevant for the future, will see its effect on a given technical indicator be suppressed amongst the indicator's values during the previous days.

Since news headlines change every day and can impact stock performance, unlike technical data which only measures the performance, it makes them a more dynamic source to cover changes in the market. Directly encoding them into model input will help to mitigate the lag of technical indicator analysis. In fact, combining sentiment analysis on news headlines with technical analysis has been shown to improve model performance for stock price prediction as opposed to solely using technical analysis \cite{6118898}.

To our knowledge, the capabilities of using Transformers in stock trend prediction have not been fully investigated past window sizes of 2 weeks \cite{transformer}. Therefore, in this work, we aim to compare our Transformer based model's performance to RNN models across longer time sequences. We experimented with window sizes of 1 week (\emph{n}=5), 2 weeks (\emph{n}=10), 3 weeks (\emph{n}=15), 5 weeks (\emph{n}=25), and 6 weeks (\emph{n}=30). Each window size is analogous to a different time-based trading strategy, with 6 weeks being the longest strategy discussed in this paper.

\section{Related Works}
Various RNN-based stock prediction methods have been applied to stock trend prediction using either technical data, sentiment data, or a combination of both, but in this paper, we focus on comparisons to the LSTM and GRU.

The LSTM\cite{Sepp} has long been used as an alternative to basic Recurrent Neural Networks for time series prediction tasks due to its ability to combat gradient vanish/explosion over longer sequences, and its ability to better capture long-term dependencies. Both of these are necessary features for a successful time-series prediction model. The model is comprised of three gates. The forget decides what information is remembered for future predictions, the input gate determines the importance of the input, and the output gate uses information passed through the previous two gates to output the next hidden state of the model. Usually, the final hidden state is used as the model's overall output. The next RNN model mentioned in our study is the GRU \cite{cho2014learning}. The GRU's approach is similar to the LSTM.  It consists of an update gate which decides on what amount of information to keep from previous time steps, and a forget gate which decides what information from previous time steps is not needed. 

Jin et al.\cite{jin2020stock} performed stock trend prediction by using an LSTM-based model to evaluate a dataset comprised of AAPL investor sentiment and technical data. The lag window was set to \emph{n}=30 meaning that their model was used to predict the opening price of the 31st day. Their model architecture, called S\_EMDAM\_LSTM, consists of a CNN component to conduct sentiment analysis followed by a final Attention-LSTM which outputted the trend prediction using the technical data and CNN sentiment output. Our work differs as our entire architecture involves only Transformer based model. 

Another study done by Khaled et al.\cite{baseline_2} was conducted using solely the LSTM architecture for stock trend prediction. This paper utilizes a Bidirectional LSTM (Bi-Dir LSTM) and a window size of \emph{n}=10 days for a prediction task similar to ours. Their dataset consisted of technical data on the S\&P 500. The implemented Bi-Dir LSTM includes two layers, where one performs operations in the normal order of the data sequence, and the other performs its operations in the reverse order of the sequence. The additional context offered by this reversed processing has been found to achieve higher performance than the Unidirectional LSTM \cite{graves2005framewise}. The model is tuned and trained on financial data but doesn't incorporate sentiment analysis.

The last work we consider is by Shahi et al.\cite{baseline_3} They conduct stock trend prediction using a dataset composed of both scraped technical and sentiment data for stocks listed in the Nepal Stock Exchange (NEPSE), Initial Public Offering (IPO), Further Public Offering (FPO). This paper uses a window size of \emph{n}=15 days to train LSTM and GRU models on the data above to conduct a comparative study on the two model types.

\section{Proposed Method}

\subsection{Dataset Description and Analysis}
Our dataset consists of 11 features: date, ticker symbol, opening price, the highest price of the day, lowest price of the day, closing price, relative strength index (RSI), exponential moving average (EMCA), simple moving average (SMA), moving average convergence divergence (MACD), and the top news headline associated with the ticker symbol and date. A ticker is a unique abbreviation used to identify a company on the stock market.

RSI, or relative strength index, indicates to an investor if a stock is being overbought or oversold. Usually if a stock is being overbought it results in a higher RSI value, and signals a good selling period. The opposite is true for oversold stocks. Measuring RSI can reveal a stock’s momentum which aids investors in predicting the magnitude and direction that a stock’s price will follow.

SMA, or simple moving average, is the average price of a stock over a certain period of time. In our case, this is the closing price. An increasing SMA value can be interpreted as an upward trend for a stock signaling investors to buy, while the opposite is true for a decreasing SMA value. EMA, or exponential moving average is similar to SMA in that it tracks the price of a stock over a period of time, however, EMA places a greater weight on more current data making it more useful for short-term trading strategies. MACD, or moving average convergence divergence, uses long-term (26 days) and short-term (12 days) EMA values to better capture the trend of a stock’s price. As previously mentioned, since EMA places a greater weight on more current data, this indicator can help introduce a broader context of market data. Since SMA, EMA, and MACD span different periods of time, using them in tandem can provide a more holistic trading strategy.

The indicator data was acquired through an EOD Historical Data API \cite{EODHD}, and the headline data is provided by the News API\cite{News}. The equations for the technical indicator data can be found below. A default window size of 50 days was used to calculate RSI, EMA, and SMA.

\begin{sloppypar}    
\begin{equation} \label{Technical Data Equation}
\begin{cases} 
      RSI = 100 - \frac{100}{1+RS}\\  
      \;\;\;\;where \; RS = \frac{Average\,Positive\,Closes}{Average\,Negative\,Closes},\\
      EMA = Close(Today) * j + EMA(Yesterday) * (1-j)\\ 
       \;\;\;\;where \;j = \frac{2}{Period_{EMA}+1},\\
      SMA = \frac{\sum_{i=1}^{SMA_{Period}}Close(i)}{SMA_{Period}},\\
      MACD = EMA_{Period = 12} - EMA_{Period = 26}\\
   \end{cases}
\end{equation}
\end{sloppypar}

Each entry in our dataset corresponds to a business day for a given stock and contains this stock's technical indicator data and top news headline for the given day. The news headline feature is preprocessed using FinBert\cite{finbert} to extract the corresponding sentiment score. We have included news headlines in our dataset since context external to the technical data which is provided by the sentiment analysis conducted on news headlines, in combination with the self-attention benefits of the Transformer architecture provides an advantage in stock trend prediction compared to conventional RNNs.

The dataset consists of 3700 entries of the FAANG (Facebook, Amazon, Apple, Netflix, Google) companies. The entries are drawn from actual stock data from a time period of close to three years spanning from 3/18/2019 - 2/18/2022. As mentioned above, the numerical data for each stock is from the EOD Historical Data API, and the news headlines are extracted from the NewsAPI. Normalization using mean and standard deviation has been seen to provide performance improvement for stock trend prediction \cite{7364089}. Therefore, we conduct a Z-Score normalization over all numerical fields to prevent the different prices of each stock in our dataset from skewing our results and to remain true to the idea of trend prediction rather than price prediction.

\subsection{Architecture Descriptions}
Our model, StockFormer, is based on the Transformer architecture. Transformers, introduced by the paper “Attention Is All You Need”\cite{attention}, are sequence-to-sequence models, that specialize in time-series analysis tasks using a concept called self-attention. When dealing with sequential data, state-of-the-art RNNs often suffer from mishandling long-term dependencies, a problem that the Transformer architecture mitigates through the use of this self-attention. To be more specific, Transformers operate by introducing this attention mechanism to the famous encoder-decoder array architecture that is commonly used in sequence translation and processing tasks. Conventionally the encoder would simply take in a sequence of data as input, but the Transformer’s attention mechanism allows the model to encode the importance of certain keywords and semantics that occur in a sentence. This then assists the decoder, which uses the encoder’s output, by giving it more contextual information.

To conduct sentiment analysis on our news headline data, we use FinBert\cite{finbert}. FinBert is the product of training Bert\cite{bert}, a multi-layer bidirectional Transformer, on a financial corpus. Bert uses WordPiece\cite{wordpiece} embeddings to encode each token in the input sequence and masks 15\% of the input tokens. This allows the model to develop a more robust bidirectional self-attention representation of the input. Training Bert on a financial corpus allows Finbert to gather sentiment scores from text with a financial perspective in mind. FinBert provides sentiment information in the form of real values that represent if the given text is positive, neutral, or negative. We use a pretrained version of Finbert to perform sentiment analysis on top news headlines in our dataset.

For some time windows, we could not find direct RNN comparisons to our prediction task that also used the same metric measurements. Therefore in this section below, we also describe a baseline LSTM we created to conduct comparisons for such time windows. This baseline shares the same sentiment analysis and technical data fusion approach with our model architecture, which allows for an accurate comparison of our model's performance and allows us to create valuable trend comparisons between LSTM-based models and StockFormer as time windows are increased.

\subsubsection{Input and Output Commonalities}
Our novel model, StockFormer, and the baseline LSTM that we describe below shares the same input format. If we are trying to predict the (\emph{n}+1) day’s normalized opening price, each model is given data on n days in the form of the features that we mentioned in the Dataset Description and Analysis section. For reference, these "raw" features include all the normalized technical indicator data along with the sentiment analysis score outputted by FinBert for each n day. Each model contains a Linear Layer for each \emph{n} day to extract raw embeddings from the raw feature data. These raw embeddings are then used differently depending on the type of model. 
When it comes to the output, there is one Linear Layer at the end of each model that takes as input the final latent output of each model and outputs the final normalized opening price prediction for the (\emph{n}+1) day. The Linear Layers in each model follow the same formula. The output is found by multiplying input elements by their respective weights. For example, given the input \emph{I} and weights \emph{W}, both of size \emph{i}, the output for a Linear Layer can be found by Equation \ref{LinearLayerEquation}:
\begin{equation} \label{LinearLayerEquation}
\sum_{i=0}^{i}W_{i}I_{i}
\end{equation}

As stated before, these layers are often implemented at the beginning of models to find intermediate embeddings by introducing nonlinearity and complexity, while also being used at the end to produce the final output.

\subsubsection{StockFormer}
In this paper, the architecture we are introducing is called the StockFormer. For this architecture, we employ the PyTorch nn.Transformer model as a backbone \cite{pytorch}. The extracted raw embeddings, which are described above, are first fed into the positional encoder portion of our model in which positional embeddings are extracted and then added to the raw embeddings to create the final input embeddings for each \emph{n} day.   

Since we are using a Transformer backbone, the \emph{n} days’ final input embeddings are not inputted sequentially, but instead in a parallel manner. This means that the position of each day with respect to others needs to be directly encoded into the final input embeddings and therefore the need to use a positional encoder arises. The positional encoder we are using is based on the sinusoidal functions (\emph{Sin}, \emph{Cos}) mentioned in the original Transformer paper \cite{attention}. These functions were both used because they are bounded and can together be represented by a linear function. Since bounded linear functions are relatively easy for neural networks to learn, these attributes make training more efficient and less intensive. Along with this, these functions can innately capture the proximity between two positions. This can be seen by the fact that the function \emph{Sin}(\emph{x}) has points closer in position to the function \emph{Sin}(\emph{x}+1) than those of \emph{Sin}(\emph{x}+2). To be more specific, the positional embedding, of the \emph{k}th token in a sequence can be found by the following formula in Equation \ref{PositionalEmbeddingEquation}:  
\begin{equation} \label{PositionalEmbeddingEquation}
 \begin{cases} 
      PE[k,2i] = Sin(\frac{k}{n^{2i/d}})\\  
      PE[k,2i + 1] = Cos(\frac{k}{n^{2i/d}})\\
   \end{cases}
\end{equation}

Here, \emph{d} is the dimension of the output positional embedding matrix (\emph{PE}) which also corresponds to the dimension of the input sequence. (\emph{PE}) contains all the positional embeddings for each token in the input sequence. In our case, each token in the input sequence corresponds to a trading day. \emph{n} is a user-defined variable that is set to 10,000 by default in the original Transformer paper \cite{attention}. Finally, the variable \emph{i} falls in the range [0, \emph{d}/2) and represents the different dimensions of the position embedding matrix which we iterate through to create the positional embedding for \emph{k}.

The final input embedding is found by adding the raw feature embeddings to the positional embeddings for each day in our input sequence. These final input embeddings are then passed into the encoder section of the Transformer backbone. In the original Transformer paper, the encoder section consists of 6 identical layers. Each layer is comprised of a multi-head attention mechanism followed by a fully connected feed-forward network. The multi-head self-attention mechanism calculates the importance of each token to the other tokens in the sequence. To create a more dynamic attention representation, multiple self-attention functions are used in this mechanic. 

We mentioned self-attention briefly before, but to be more specific, self-attention can be described as using learnable parameters and dot product operations to query each token in a sequence against the entire sequence to determine a corresponding reweighting factor. This reweighting factor is then used on the sequence to derive an embedding for the queried token with a greater context of its relations to the remaining tokens in the sequence. 

To conduct the self-attention function, first select a token to reweight the embedding for. The chosen token is passed through a Linear Network to derive its query embedding, while all the tokens in the sequence are passed through a separate Linear Network to determine their key embeddings. Then conduct a dot product of the query embedding against the key embeddings and perform a SoftMax operation on the result to derive the final reweighting factor, which consists of one corresponding reweighting scalar for each value embedding. Value embeddings are found by passing all the sequence tokens through another Linear Network, similar to how key embeddings are found. Finally, multiply the value embeddings of the sequence by their corresponding scalar values contained in the reweighting factor and sum the results to derive the final reweighted embedding of the token associated with the query. In this manner, by conducting self-attention on our input sequence of \emph{n} days, we can factor in the importance of each day in a sequence to the others by reweighting their embeddings which will give us better context when predicting the (\emph{n}+1) day’s normalized opening price. Each previously mentioned function in the encoder's multi-head self-attention attention mechanism uses the following formula in Equation \ref{AttentionEquation}:

\begin{equation} \label{AttentionEquation}
Attention(Q, K , V) = \frac{QK^T}{d} V
\end{equation}

Here, \emph{d} is the dimension of the input. The aforementioned query, key, and value embeddings are represented as matrices to parallelize the self-attention process for all tokens in the sequence. \emph{Q} is the query matrix, \emph{K} is the key matrix, and \emph{V} is the value matrix. As a reminder, these matrices can be found by conducting the dot product of their corresponding Linear Layer (weighted matrices) by the input tokens. For example, \emph{Q} is the result of conducting the dot product between the query inputs and the weighted matrix associated with query inputs. The final output of the model encoder, which is a feed-forward Linear Network that uses these new attention embeddings as input, is a sequence of continuous representations that are used by the decoder to extract the final intermediate embedding corresponding to the (\emph{n}+1) day’s opening price. 

Similarly to the encoder, the decoder consists of 6 identical layers containing similar elements. Each layer contains a multi-headed attention mechanism that performs self-attention on the decoder's input. In our case, the decoder's input consists of the encoder's output along with the positionally encoded normalized opening prices of the previous \emph{n} days. The outputs of the self-attention operations performed on both portions of the decoder's input are then passed into a fully connected feed-forward network which outputs a final shifted intermediate representation of the normalized open prices for days [2, \emph{n}+1]. As mentioned in the Input and Output Commonalities subsection, this intermediate representation is passed through a final Linear Layer whose output represents the (\emph{n}+1) day's normalized opening price.

\begin{figure*}[]
    \centering
    \includegraphics[width=\textwidth]{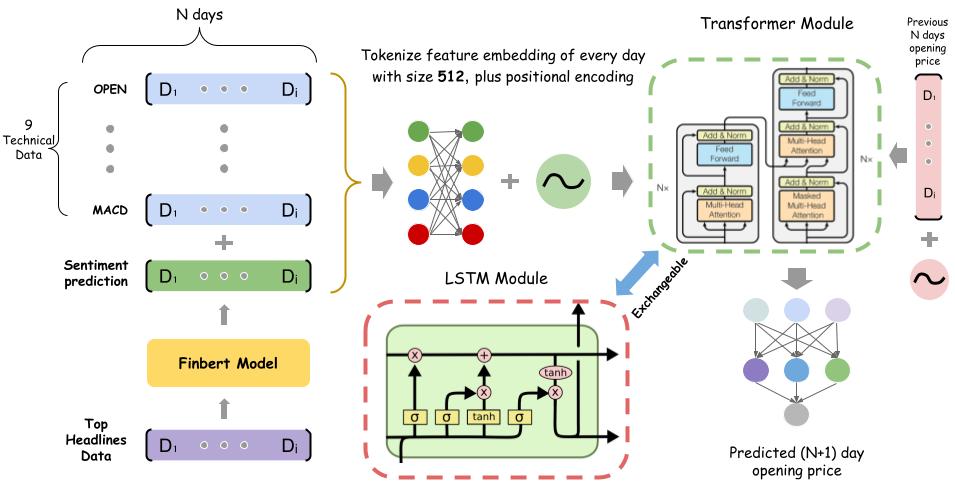}
    \caption{StockFormer WorkFlow with Exchangeable Baseline LSTM Module}
    \label{fig: Stockformer WorkFlow}
\end{figure*}

The complete workflow of this model, beginning from initial feature latent embedding creation to final Transformer output, can be seen in Figure \ref{fig: Stockformer WorkFlow}.

\subsubsection{LSTM Baseline}
The model that we chose to be our baseline for time windows that did not have readily available RNN models for comparison, was the LSTM. Our LSTM model uses the PyTorch nn.LSTM as a backbone \cite{pytorch}. Since this is a recurrent neural network, there is no need for the Positional Encoder as the raw embeddings for each \emph{n} day are inputted sequentially. In this model, the final hidden state is used as input for the final Linear Layer which outputs the final normalized opening price prediction for the (\emph{n}+1) day.

To get the best performance from our baseline LSTM, we made it bidirectional. This meant that instead of information flowing from beginning to end, it also went from end to beginning. Therefore this preservation of information from both the past and future can help add complexity to the model and allow it to better represent the data. We also found that the LSTM performed the best when hidden dimension values were set to 10 and when we stacked two LSTMs together to create a double-layered bidirectional LSTM. The workflow for this baseline model using the LSTM module can be seen in Figure \ref{fig: Stockformer WorkFlow} as well.


\section{Experiment and Results Analysis}
Since our problem statement is centered around using \emph{n} days to predict the (\emph{n}+1) day’s opening price, we decided to structure our experiments in a format that varied the value of \emph{n}. Varying this so-called window of days used for prediction was done to try and compare the different models’ abilities to capture long and short-term dependencies.

\subsection{Metrics}
We used four major metrics to measure the performance of each model. For loss measurement, MSE loss was used due to the fact that predicting the (\emph{n}+1) day’s open price is a regression-based task. MSE can be measured by taking the average of the difference between each data entry’s corresponding ground truth label and predicted model output. The formula for MSE loss can be found below in Equation \ref{MSEEquation}. For this and the following formulas, \emph{N} is the total amount of data points, \emph{x} is the ground truth open price data, and \emph{y} is the predicted open price data:

\begin{equation} \label{MSEEquation}
\frac{1}{N}\sum_{i=1}^{n}(x_i-y_i)^2
\end{equation}

Along with MSE Loss, R2 score was also used. R2 score is a metric that is specifically made to measure the performance of regression models with a higher R2 score being considered better. R2 score essentially tells you how well your model fits the train/test data's regression line by taking into account the residuals, which are the distances between each predicted value and its corresponding ground truth value. The formula for R2 is shown below in Equation \ref{R2Equation}:
\begin{equation} \label{R2Equation}
1-\sum_{i=1}^{N}\frac{(x_i-y_i)^2}{(x_i-\bar{x})}
\end{equation}

The numerator of the summation represents the sum squared regression which is the sum of the residuals squared. The denominator of the summation represents the total sum of squares which squares the distance between the ground truth data and the mean of the total ground truth data.

The next metric that we used was AUC score. Normally AUC measures the effectiveness of classification models, but we augmented it to fit our regression task. The AUC score of a classifier is equal to the probability that the classifier gives a higher score to a positive example compared to a negative example. Therefore a regression AUC score in the context of our problem could be obtained in a similar fashion; if we take any two ground truth observations \emph{a} and \emph{b} such that \emph{a}’s true open price is greater than \emph{b}’s true open price, then the regression AUC score is equal to the probability that our model actually ranks \emph{a}’s predicted open price higher than \emph{b}’s predicted open price. In this case, a higher AUC score would still be more desirable. Therefore for the number of pairs \emph{P} described above, containing \emph{a} and \emph{b}, the AUC score can be found using the formula in Equation \ref{AUCEquation} below:

\begin{equation} \label{AUCEquation}
 \frac{1}{P}\sum_{i=1}^{P} \begin{cases} 
      predicted_a > predicted_b = 1\\  
      predicted_a < predicted_b = 0\\ 
      predicted_a == predicted_b = 0.5\\ 
   \end{cases}
\end{equation}

The final metric we used for measurement was directional accuracy, which can be seen in Equation \ref{DAEquation}. This measure is often used by regression-based stock prediction models as a convenient way to measure their accuracy. In our context, the directional accuracy of a model can be found by checking if the (\emph{n}+1) day's ground truth open price increases or decreases in relation to the \emph{n} day's ground truth open price. If the (\emph{n}+1) day's ground truth price and predicted price both increase or both decrease in relation to the n day's ground truth open price, then the directional accuracy increases. The directional accuracy also increases if both the (\emph{n}+1) days open price ground truth and prediction are equal to the \emph{n} day's ground truth open price.

\begin{equation} \label{DAEquation}
     \frac{1}{N}\sum_{i=1}^{n} \begin{cases} 
          (x_i >= x_{i-1}) \land (y_i >=  x_{i-1}) = 1\\  
          (x_i < x_{i-1}) \land (y_i <  x_{i-1}) = 1\\ 
          else = 0\\ 
       \end{cases}
\end{equation}

It is important to note that the LSTM models for which we draw comparisons as baselines do not measure performance using every metric that we use to measure the performance of our StockFormer model. Therefore our experimental analysis is unique for each LSTM model and its associated time window. However, even for baselines that contained missing metrics we could not compare to, we still chose to display all StockFormer metric measurements as this reveals valuable trends in performance. 

\subsection{Best Run Parameters}
Some training parameters for the StockFormer were consistent amongst all our runs. For each time window, the model was trained for 50 epochs. Along with this, we used the Adam optimizer for all runs due to its ability to converge quickly. Finally, a learning rate of 0.0001 consistently led to the best convergence across all experiments.

\subsection{Best Model Parameters}
For our StockFormer model, the best performance across all-time windows came when we had 6 encoder and decoder layers stacked on top of each other. The hidden dimension of each layer’s input and output was set to 80 as well. We also incorporated a dropout value of 0.1 which we believe helped the model generalize better. Finally, the number of attention heads that we found to be most beneficial was 8. Each attention head has its own set of weights that are used when self-attention is calculated by the Transformer backbone. Intuitively, it is better to have multiple attention heads as this gives the model more opportunity to learn different attention representations that can be useful when conducting the final prediction. 

\subsection{Experimental Results}
\begin{figure}
    \begin{minipage}[c]{1\linewidth}
        \includegraphics[width=\linewidth]{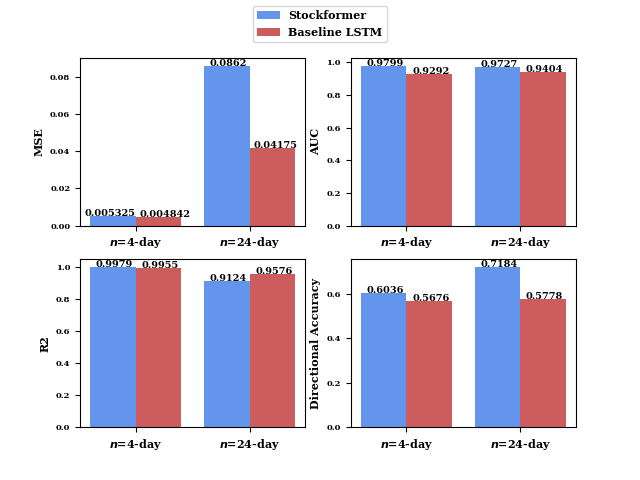}
        \caption{Baseline LSTM Performance Comparison}
        \label{fig:Baseline LSTM Performance Comparison}
    \end{minipage}%
    \hfill
    \begin{minipage}[c]{1\linewidth}
        \includegraphics[width=\linewidth]{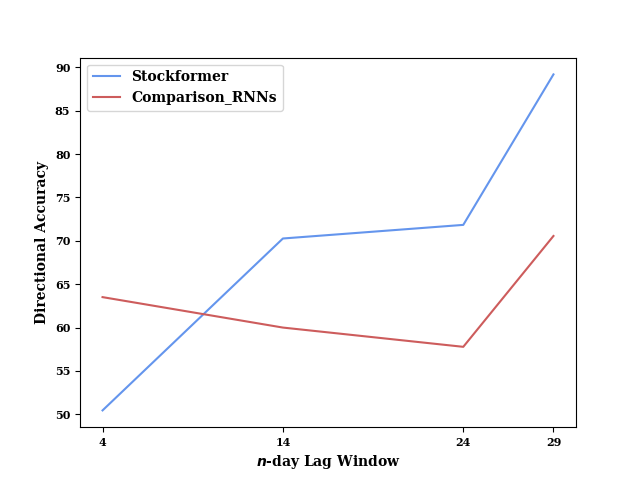}
        \caption{Accuracy Trend Comparison}
        \label{fig:Accuracy Trend Comparison}
    \end{minipage}

\end{figure}

As a reminder, this baseline LSTM was created as an evaluation method for time windows that we could not find comparable studies on. For a lag period of \emph{n}=4 days, a difference of 3.6\% directional accuracy reveals that our model performs better than the baseline LSTM on short time ranges. As we increase the lag period to \emph{n}=24 days, we can see that our model reaches an AUC score of 0.9727 and directional accuracy of 0.7184 as it continues to outperform our baseline LSTM while the differences in loss remain minimal. The full comparison of our model's performance, in blue, with respect to our baseline LSTM's performance, in red, can be seen in Figure \ref{fig:Baseline LSTM Performance Comparison}. 

This trend of outperforming RNNs as the lag period increases is visible even as we compare it to the models mentioned in our Related Works Section. We present our model's performance to these models in Table \ref{fig:Related Model Performance Comparison}. The Bi-Dir LSTM that we use to conduct a comparison for a lag period of \emph{n}=9 days did not contain an accuracy measurement in its paper. However, we can see that our model's performance stayed similar as we increased the lag period from \emph{n}=4 to \emph{n}=9 days which is the first indication of our model's ability to better capture long-term dependencies. As we move to a lag period of \emph{n}=14 days, we see that our model completely outperforms the comparison RNNS. Again, cannot compare loss measurements here because the LSTM-News and GRU-News models were not trained on normalized data. That being said, the increase in directional accuracy to 0.7027 is enough to support a positive correlation between increased window size and our model’s performance. Finally, the greatest performance in directional accuracy of our model can be found if the lag period is set to \emph{n}=29 days. The model we are comparing here is the S\_EMDAM\_LSTM model mentioned earlier. Our model was able to reach 0.8919 in directional accuracy compared to 0.7056 for S\_EMDAM\_LSTM. This is the largest difference in accuracy that we found.

Overall, is clear that our model does not perform as well for a time window of 5 days compared to the rest. We believe this is attributed to the fact that a lag period of 4 days is not long enough for our Transformer based model's ability to better extract long-term dependencies to become an advantage. However, it is important to note that our model's performance improved overall as the sequence length increased, even when compared to RNN baselines that also incorporate sentiment analysis. This is evident when comparing the directional accuracy trends of our StockFormer and the aforementioned RNN baselines which can be seen in Figure \ref{fig:Accuracy Trend Comparison}. We believe these results were caused by the StockFormer being able to better keep track of long-range dependencies using self-attention due to its Transformer backbone. Even if we were unable to compare all metrics to each baseline, identifying this trend is useful as it supports the narrative that Transformers can capture long-term dependencies better than RNNs in the financial domain. As we stated before, longer stock prediction time windows with Transformers have not been explored as much to our knowledge, so this body of work serves as an incentive to continue this area of research further.



{\Huge
\begin{table*}[]
\centering
\resizebox{\textwidth}{!}{%
\begin{tabular}{|c|ccc|ccc|ccc|ccc|ccc|}
\hline
           & \multicolumn{3}{c|}{StockFormer}                                                                                                             & \multicolumn{3}{c|}{Bi-Dir LSTM\cite{baseline_2}}                                                                                                      & \multicolumn{3}{c|}{LSTM-News\cite{baseline_3}}                                                                                              & \multicolumn{3}{c|}{GRU-News\cite{baseline_3}}                                                                                               & \multicolumn{3}{c|}{S\_EMDAM\_LSTM\cite{{jin2020stock}} }                                                                                                           \\ \hline
Lag Period & \multicolumn{1}{c|}{MSE Loss}        & \multicolumn{1}{c|}{R2}              & \begin{tabular}[c]{@{}c@{}}Directional\\ Accuracy\end{tabular} & \multicolumn{1}{c|}{MSE Loss}           & \multicolumn{1}{c|}{R2}    & \begin{tabular}[c]{@{}c@{}}Directional\\ Accuracy\end{tabular} & \multicolumn{1}{c|}{MSE Loss} & \multicolumn{1}{c|}{R2}    & \begin{tabular}[c]{@{}c@{}}Directional\\ Accuracy\end{tabular} & \multicolumn{1}{c|}{MSE Loss} & \multicolumn{1}{c|}{R2}    & \begin{tabular}[c]{@{}c@{}}Directional\\ Accuracy\end{tabular} & \multicolumn{1}{c|}{MSE Loss}      & \multicolumn{1}{c|}{R2}                & \begin{tabular}[c]{@{}c@{}}Directional\\ Accuracy\end{tabular} \\ \hline
n=9     & \multicolumn{1}{c|}{0.004659}        & \multicolumn{1}{c|}{\textbf{0.9956}} & 0.5766                                                         & \multicolumn{1}{c|}{\textbf{0.0009356}} & \multicolumn{1}{c|}{0.994} & --                                                             & \multicolumn{1}{c|}{--}       & \multicolumn{1}{c|}{--}    & --                                                             & \multicolumn{1}{c|}{--}       & \multicolumn{1}{c|}{--}    & --                                                             & \multicolumn{1}{c|}{--}            & \multicolumn{1}{c|}{--}                & --                                                             \\ \hline
n=14     & \multicolumn{1}{c|}{0.003411}        & \multicolumn{1}{c|}{\textbf{0.9968}} & \textbf{0.7027}                                                & \multicolumn{1}{c|}{--}                 & \multicolumn{1}{c|}{--}    & --                                                             & \multicolumn{1}{c|}{4.8031}   & \multicolumn{1}{c|}{0.979} & 0.60                                                           & \multicolumn{1}{c|}{5.399}    & \multicolumn{1}{c|}{0.967} & 0.59                                                           & \multicolumn{1}{c|}{--}            & \multicolumn{1}{c|}{--}                & --                                                             \\ \hline
n=29     & \multicolumn{1}{c|}{0.1071} & \multicolumn{1}{c|}{0.9219}          & \textbf{0.8919}                                                & \multicolumn{1}{c|}{--}                 & \multicolumn{1}{c|}{--}    & --                                                             & \multicolumn{1}{c|}{--}       & \multicolumn{1}{c|}{--}    & --                                                             & \multicolumn{1}{c|}{--}       & \multicolumn{1}{c|}{--}    & --                                                             & \multicolumn{1}{c|}{10.2178296132} & \multicolumn{1}{c|}{\textbf{0.977388}} & 0.7056                                                         \\ \hline
\end{tabular}%
}
\caption{Related Model Performance Comparison}
\label{fig:Related Model Performance Comparison}
\end{table*}
}

\subsection{Comparison to Recent Transformer Based Stock Prediction Model}
To complete the reporting of our results, we figured it was appropriate to mention a recent paper, that used a Transformer as a backbone for its initial feature extraction in order to conduct stock trend prediction. While this paper is also an example of a recent Transformer based stock prediction paper that does not explore longer time windows, it can still serve as an important point of reference for the short-term capabilities of our model. The authors of this paper decided to combine a Transformer and LSTM in order to create a model that would conduct stock prediction using financial and news over a 5 day window \cite{transformer}. This model, called TEANet, first extracts intermediate features from its technical stock and tweet text data using a Transformer encoder-based architecture. It then uses these intermediate features as input to its LSTM to extract temporal features which represent the fusion of stock and tweet data. This fusion is then used as input into the temporal attention portion of the model. When trained on a dataset similar to ours that used Twitter tweets rather than news headlines, this model achieved an accuracy of 65.16\% over a 5 day period. As shown in Figure \ref{fig:Baseline LSTM Performance Comparison}, our StockFormer managed an accuracy of 60.36\% over a similar period of time (\emph{n}+1=5 day). We believe this performance discrepancy comes from the increased complexity that TEANet offers by using an LSTM, which we have shown performs similar to our Transformer based architecture over this smaller 5 day period. While it may be useful to incorporate an RNN in a similar manner into our StockFormer architecture, we would also like to see how TEANet would perform on longer time windows, such as \emph{n}=14 days and beyond. We have seen that an LSTM's performance can degrade during this time period, so it would be interesting to see if an LSTM inclusive architecture would be affected by this trend as well.

\section{Conclusion}
Based on the results of our experiments, the combination of sentiment analysis and Transformers is a viable approach for stock trend prediction. Testing our Transformer implementation on varying time windows displays the model's ability to better capture long-term dependencies than comparable RNNs. This better performance over longer time windows also indicates the model's usefulness for long-term based trading strategies. In the future, we aim to further develop this concept through a few modifications. Our longest time window in this paper consists of 6 weeks, so testing time windows of greater lengths could indicate even better comparative performance to RNN models. Along with this, expanding the number of tickers tested and incorporating new technical indicator features for global market data may also increase the performance of the model. An alternate task we could try is switching from a regression problem to a binary classification problem, by predicting strictly if a given stock's price increases or decreases. This type of problem formulation allows us to diversify the metrics available for evaluating the model as well. More technical changes also include experimenting with different positional encoding functions. Currently, the model uses a sinusoidal function for this purpose, but it is possible a more optimal function or even a trained time-series positional encoder exists that can be better tuned for stock time-series prediction. Finally, the raw encoding function can be made more sophisticated as well. Instead of using a single linear layer for each preceding day, we can perhaps incorporate an RNN or a more advanced Neural Network to extract such embeddings to add complexity to our model. 

\bibliographystyle{IEEEtran}
\bibliography{references.bib}
\end{sloppypar}
\end{document}